\documentclass[twocolumn,showpacs,preprintnumbers,amsmath,amssymb,prb]{revtex4-1}
\usepackage{graphicx}
\usepackage{amsmath, amsthm, amssymb}
\usepackage{latexsym}
\usepackage{amssymb}
\usepackage{bm}
\usepackage{dcolumn}
\usepackage{epstopdf}
\usepackage{color}

\begin{document}

\title{Spin-pairing and penetration depth measurements from nuclear magnetic resonance in NaFe$_{0.975}$Co$_{0.025}$As}

\author{Sangwon Oh$^{1}$, A. M. Mounce$^{1}$, Jeongseop A. Lee$^{1}$, W. P. Halperin$^{1}$, C. L. Zhang$^{2}$, S. Carr$^{2}$, Pengcheng Dai$^{2}$}

\affiliation{$^1$Department of Physics and Astronomy, Northwestern University, Evanston, Illinois 60208, USA \\
$^2$Department of Physics and Astronomy, The University of Tennessee, Knoxville,
Tennessee 37996, USA}
\date{Version \today}

\begin{abstract}
We have performed $^{75}$As nuclear magnetic resonance (NMR) Knight shift measurements on single crystals of NaFe$_{0.975}$Co$_{0.025}$As to show that its superconductivity is a spin-paired, singlet state consistent with predictions of the weak-coupling BCS theory.  We use a spectator nucleus, $^{23}$Na, uncoupled from the superconducting condensate, to  determine the diamagnetic magnetization and to correct for its effect on the  $^{75}$As NMR spectra. The resulting temperature dependence of the spin susceptibility follows the Yosida function as predicted by BCS for an isotropic, single-valued energy gap.  Additionally, we have analyzed the $^{23}$Na spectra that become significantly broadened by vortices to obtain the superconducting penetration depth as a function of  temperature with $\lambda_{ab}(0) = 5,327 \pm$ 78$\,\AA$.
\end{abstract} 

\pacs{ }

\maketitle

\section{INTRODUCTION}
There have been many remarkable discoveries of  superconducting compounds that are almost antiferromagnetic. Some have unconventional broken symmetries with very high transition temperatures such as with the cuprates or with multiple vortex phases like the heavy fermion compound UPt$_3$.~\cite{nor13} The recently discovered iron-pnictide family is in a new class which appear to be conventional but with an interplay with magnetism that is not well understood.  An important question in all of these cases is whether an appropriate theoretical starting point is the classic work of Bardeen, Cooper, and Schrieffer~\cite{bar57} (BCS) where electrons conventionally form Cooper pairs of opposite spin and angular momenta, weakly coupled together.  Unconventional Cooper pairing was first found in liquid $^3$He~\cite{lee96} where attraction between fermions originates from  spin-fluctuations in a nearly-magnetic Fermi liquid, a rather different mechanism for superconductivity than proposed by BCS .  Nonetheless, it is important to keep in mind that the BCS weak-coupling framework accurately accounts for the thermodynamics of this superfluid including its spin susceptibility.  The proximity of superconductivity and antiferromagnetism in pnictide materials is similarly suggestive, but a comparison to BCS theory has not yet been established.  These superconductors are inherently multi-band with different pairing strengths and energy gaps on hole and electron portions of their Fermi surface, complicating the interpretation of experiments.

However, for one of the pnictide family, NaFe$_{1-x}$Co$_{x}$As, angle-resolved photo-emission spectroscopy (ARPES)~\cite{liu11,thi12}, $x$\,=\,0.050, and scanning tunneling microscopy (STM) measurements~\cite{cai13,yan12,wan13}, $x$\,=\,0.025,\,0.050, indicate that it has a single energy gap or two gaps of the same size.  In this case a thermodynamic analysis of the temperature dependence of the spin susceptibility is straightforward.  We have performed $^{75}$As nuclear magnetic resonance (NMR) Knight shift measurements of the spin susceptibility on clean crystals of NaFe$_{0.975}$Co$_{0.025}$As (NaCo25) to show that opposite spins pair in a singlet state, quantitatively following the temperature dependence expected from the weak-coupling BCS theory expressed by the Yosida function. An essential step in our procedure to extract the spin part of the Knight shift  is  to measure  the local diamagnetic fields from $^{23}$Na NMR and subtract them from the $^{75}$As spectra.

\section{EXPERIMENTAL METHODS} 
Our $^{75}$As and $^{23}$Na NMR experiments were performed at Northwestern University  from 4.2 K to room temperature with external magnetic field $H=6.4$ T parallel to the $c$-axis. Because of the high reactivity of Na we used an environmentally sealed sample holder, Fig.1(a),  made out of glass or Stycast 1266$\textsuperscript{\textregistered}$. The containers  were filled with helium in an oxygen free chamber  to prevent sample degradation and to exchange heat.  Degradation can be identified over time from an increase in the NMR linewidth and consequently several different crystals were used.  In the work we report here, both  nuclei exhibited very narrow NMR spectra at room temperature:  8 kHz for $^{75}$As, and 3 kHz for $^{23}$Na.  Crystals with dimensions $\sim 3 \times 2 \times 0.3$ mm$^{3}$ were grown at the University of Tennessee and found to have $T_c$ of 21 K from magnetization with $H=10$ G parallel to the $ab$-plane, Fig.1(b).  Spin-echo sequences ($\pi/2$ - $\pi$) were used to obtain spectra, Knight shift, and spin-lattice relaxation rate, $1/T_1$, from the central transition (-1/2 $\leftrightarrow$1/2) with a $\pi$-pulse $\approx$ 8 $\mu$sec. The spin-spin relaxation time, $T_2$, was measured with a Carr-Purcell-Meiboom-Gill (CPMG) sequence and $T_1$ was obtained with a saturation recovery method for $T_1<2$ sec. For longer $T_1$ the more efficient progressive saturation technique~\cite{mit01a} was used.

\section{Results and Discussion}
\begin{figure}[!ht]
\centerline{\includegraphics[width=0.5\textwidth]{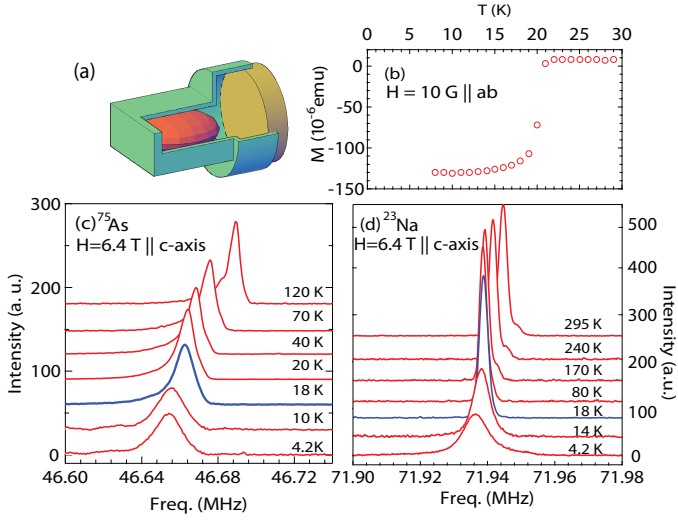}}
\caption{(a) Schematic drawing of the sample holder; (b)  magnetization measurements; (c, d) NMR spectra of NaCo25 in 6.4 T with H $||$ c-axis. The narrow linewidths confirm high sample quality. As with other pnictide superconductors~\cite{oh11,oh12,nin10} NaCo25 shows a continuous decrease on cooling from room temperature above $T_c$.  Below $T_c$ = 18 K (blue trace) the spectra shift noticeably to lower frequency with onset of superconductivity. The increased linewidth on cooling below $T_c$ is due to the formation of a vortex lattice.}

\label{fig1}
\end{figure}

The $^{75}$As and $^{23}$Na NMR spectra are shown in Fig.1(c) and (d). From room temperature to just above the transition temperature, $T_c = 18$ K in 6.4 T, the linewidths of the spectra are very narrow, less than 10 kHz, and have very weak dependence on temperature, a clear indication of high crystal quality.  On cooling from high temperatures in the normal state,  both $^{75}$As and $^{23}$Na spectra shift to  lower frequency as has been observed in other pnictide superconductors.~\cite{oh11,oh12,nin10}  The superconducting transition at $T_c$, is evident in an abrupt decrease in the peak frequency of the spectra, Fig.2.  Formation of the vortex lattice generates an inhomogeneous distribution of local fields, broadening both spectra in the superconducting state.

The  measured NMR frequency shift, $\Delta\omega(T)$,  relative to the Larmor frequency, $\Delta\omega(T)/ \omega_L =  (\omega(T) -\omega_L)/\omega_L$, can be expressed as,
\begin{equation}
 K(T)= K_{s}(T) +K_{orb}+K_{Q}+4\pi(1-D) M(T)/H,
\end{equation}
\noindent

\begin{figure}[!ht]
\centerline{\includegraphics[width=0.45\textwidth]{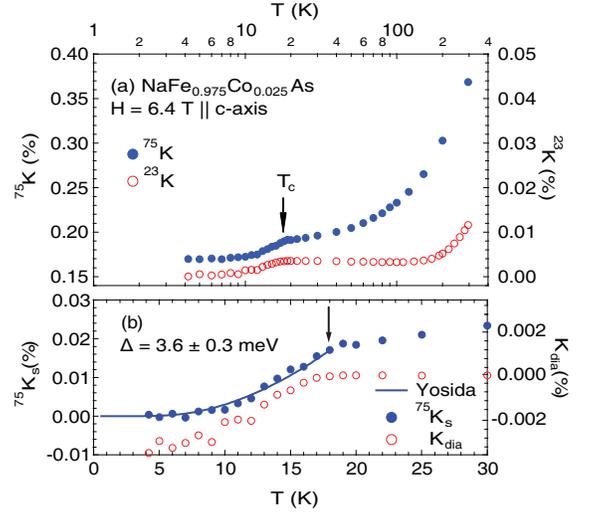}}
\caption{Knight shifts for $^{75}$As and $^{23}$Na over a wide temperature range.  (a) The  decrease of $K$ in the normal state on cooling is characteristic of pnictide superconductors. (b) The low temperature behavior of $^{75}K_s$, after subtraction of $K_{dia}$, is fitted to a Yosida function,~\cite{yos58} solid curve, with $2\Delta(0)$/$k_B T_c$ = 4.0  which we interpret as evidence for  weak-coupling superconductivity.}
\label{fig2}
\end{figure}

generically referred to as the Knight shift, $^{75}K$ and $^{23}K$ for each nucleus,  determined from the peak positions of the spectra in Fig.2. The Knight shift has electron spin, $K_s$, orbital, $K_{orb}$, quadrupolar, $K_{Q}$, and $K_{dia}$ terms, the latter from the superconducting diamagnetism, $M(T)$ with demagnetization factor, $D$; $K_{orb}$ and $K_{Q}$ provide a temperature independent offset. In the normal state the temperature dependence comes from $K_s(T) = \chi_s(T) A_{hf}$, given by the spin susceptibility, $\chi_s(T)$, and its corresponding hyperfine coupling, $A_{hf}$.  For $T> 100$ K the different temperature  dependence of $^{75}K$ and $^{23}K$ is due to a much smaller hyperfine field for Na as compared to As, which we estimate to be $^{75}A_{hf}/^{23}A_{hf} \approx 19$. For $T < 100$ K the comparison of $^{23}K$ with $^{75}K$ indicates $^{23}K_s$ is negligible and $^{23}K_{orb}+^{23}K_{Q} = 0.0033\,\%$.  The decrease of $^{23}K$ in the superconducting state is due entirely to $K_{dia}(T)$ and provides an internal measure of the local field that can be subtracted from $^{75}K$ to give a precise determination of the spin shift, $^{75}K_s(T<T_c)$, Fig.2(b), proportional to the spin susceptibility.

Based on ARPES~\cite{liu11,thi12} and STM~\cite{cai13,yan12,wan13} experiments we infer that there is a single isotropic superconducting energy gap in NaCo25. Even though the sizes of the superconducting gaps from different ARPES measurements in NaFe$_{0.95}$Co$_{0.05}$As (NaCo50) are different, they are consistent in two ways: the energy gaps in NaCo50 are isotropic, and the sizes of the gaps on the electron and hole Fermi surfaces are almost identical.~\cite{thi12,liu11} Consequently, we fit the temperature dependence of $^{75}K_s$ to an isotropic Yosida function, $Y(T)$,~\cite{yos58} Eq.2, to determine the energy gap, $\Delta(T)$.  Allowing for strong coupling, we take $\Delta(T)$ to be scaled~\cite{pad73} by a factor $\alpha$ from the weak-coupling BCS energy gap, $\Delta_0(T)$, represented in Eq.3 by a convenient analytical form,~\cite{hal90} and then we fit the measured Knight shift to determine $\alpha$  and hence $\Delta(T)$.

\begin{figure}[!ht]
\centerline{\includegraphics[width=0.45\textwidth]{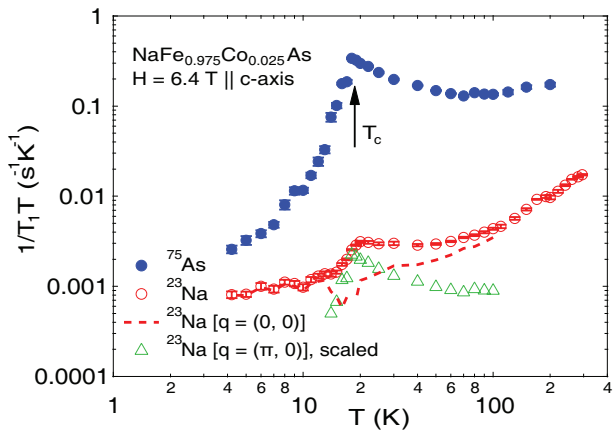}}
\caption{The spin-lattice relaxation rate divided by temperature, $1/T_1T$, from $^{75}$As and $^{23}$Na. The increase of $1/^{75}T_1T$ (blue circles) on cooling in the normal state is due to spin fluctuations,~\cite{nin10} and  drops for $T< T_c$ with the onset of superconductivity. Despite the small hyperfine field, $1/^{23}T_1T$, has a similar behavior near $T_c$ (red circles).  Scaling the rates by the square of the ratio of the hyperfine fields and their respective gyromagnetic ratios gives the open green triangles for the expected contribution at the Na site.  What remains is a second component of relaxation (red dashed curve) of unknown origin.}
\label{fig3}
\end{figure}

\noindent
For $T<T_c$,
\begin{align}
\label{eq1}
\frac{K_s(T)}{K_s(T_c)} &= Y(T) = 1-2\pi k_B T \sum_{n=0}^{\infty}\frac{\Delta(T)^2}{(\epsilon_n^2+\Delta(T)^2)^{3/2}}\\
\Delta(T)&=\alpha \Delta_{0}(T)\\
\Delta_{0}(T) &=\Delta_{0}(0)\tanh(\dfrac{k_{B}T_{c}\pi}{\Delta_{0}(0)}\sqrt{\dfrac{2}{3}(\dfrac{T_c}{T}-1)\left.\dfrac{\Delta C}{C}\right\vert_{0}})
\end{align} 
\noindent
where $\epsilon_n= 2\pi(n+1/2)k_BT$~\cite{sau12} and $\frac{\Delta C}{C}\vert_0$ is 1.43.

With this procedure we find the superconducting gap size to be $3.6 \pm 0.3$ meV, $2\Delta/k_{B}T_{c}$ = 4.0, and $\alpha = 1.14\pm 0.1$. The resultant fit is shown in Fig.2.(b). Our result is slightly larger than the BCS limit and supports the basic idea that  superconductivity in NaCo25 is weakly-coupled consistent with a recent ARPES measurement.~\cite{thi12}  STM measurements~\cite{wan13} on the same material and from the same source as ours, reports a larger energy gap of 5.5 meV. It is not clear what is the origin of this difference except that one measurement is for the charge channel and the other for spin.  It is interesting anecdotally that the bosonic mode reported in that work is close in energy, within 8\%,  to $2\Delta$ inferred from the spin-Knight shift.

We also report our measurements of spin-lattice relaxation rates for both $^{75}$As and $^{23}$Na NMR and the evidence they provide for spin fluctuations in the normal and superconducting states. Importantly, the comparison between the rates for the two nuclei confirm the relative magnitude of the hyperfine fields we deduced from Knight shifts at high temperatures.  This supports our claim that $^{23}$Na is largely uncoupled from superconductivity and can serve as a spectator nucleus for the measurement of diamagnetic local fields.

\begin{figure}[!ht]
\centerline{\includegraphics[width=0.5\textwidth]{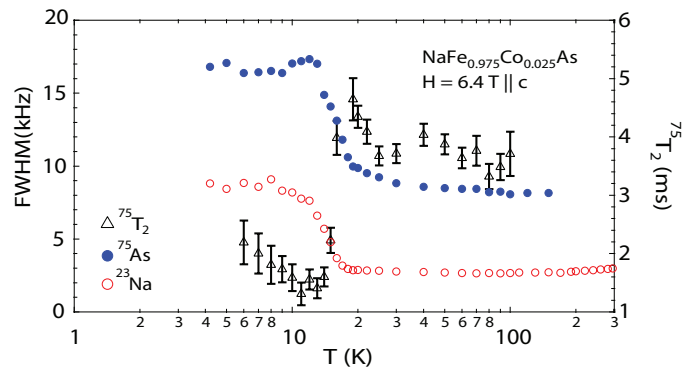}}
\caption { Temperature dependence of the linewidth, full-width at half-maximum (FWHM), of the $^{75}$As and $^{23}$Na central transition, and the spin-spin relaxation time, $T_2$. The narrow linewidths at $T=20$ K from $^{75}$As (10 kHz) and $^{23}$Na (3 kHz) and the weak temperature dependences indicate high crystal quality.  Below $T_{c}$ the linewidth increases in the   inhomogeneous vortex state and becomes constant below $\approx 13$ K for $^{75}$As (12 K for $^{23}$Na).}
\label{fig4}
\end{figure}

In contrast to the Knight shift ($\omega\approx 0, q \approx 0$) the spin-lattice relaxation is determined by the sum over wave vectors, $q$, of the imaginary part of the dynamic susceptibility which is  coupled to the nucleus through the square of the hyperfine field given by the form factor, $F(q)$, Eq.5,
\begin{eqnarray}
\label{eq2}
1/T_{1}T \propto \gamma^2\sum_q{F(q)}\dfrac{Im [\chi_{\bot}(q,\omega)]}{\hbar\omega}\ 
\end{eqnarray} 
\noindent
where $\gamma$ is  the gyromagnetic ratio, $F(q)$ the form factor, Im [$\chi_{\bot}$] is the imaginary part of the dynamic susceptibility perpendicular to the field.

We have found different behavior for $1/^{75}T_1T$ and $1/^{23}T_1T$ , Fig.3. In the normal state, $1/^{75}T_1T$ increases with decreasing temperature approaching $T_c$  and can be understood in terms of spin fluctuations~\cite{nin10,kit11,li11,oh12} expressed in terms of the Moriya theory,~\cite{mor63} Eq.5.The increase is suppressed with the onset of superconductivity, so that $1/^{75}T_1T$ drops dramatically below $T_c$, apparently strongly dependent on the density of normal quasiparticle excitations.  For Na we find that $1/^{23}T_1T$ has a component that mimics this behavior and can be accounted for by the weak hyperfine coupling strength, $^{23}A_{hf}$. Scaling the rates by the square of the ratio of the hyperfine fields and their  respective gyromagnetic ratios, gives the open green triangles in Fig.3.  Subtracting this spin-fluctuation component of $1/^{23}T_1T$ leaves a second contribution represented by the dashed red curve. We speculate that the spin-fluctuations come from q $\approx$ ($\pi$,0) or (0,$\pi$ ) and that the remaining contribution might be from, q $\approx$ (0,0) not significantly affected by superconductivity, but of unknown origin.

\begin{figure}[!ht]
\centerline{\includegraphics[width=0.45\textwidth]{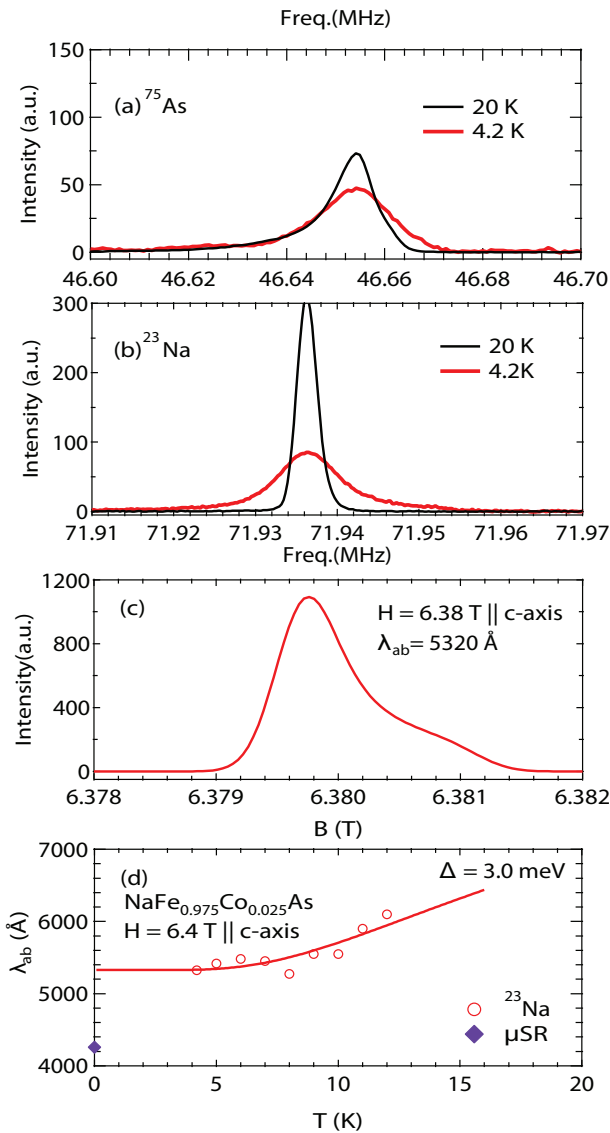}}
\caption {Spectra in the normal state (20 K, shifted in frequency for comparison) and the superconducting state (4.2 K) for (a)$^{75}$As and (b)$^{23}$Na. The linewidth broadening due to vortices is evident in the superconducting state, especially clear for $^{23}$Na. (c) The Ginzburg-Landau calculation that best fits the superconducting state $^{23}$Na spectra at $T = 0$ K from which the  penetration depth, $\lambda_{ab}$, was determined. (d) Penetration depth as a function of temperature where $\lambda_{ab}(0)$ is $5,327\pm78\,\AA$. Temperature dependence of the penetration depth is shown in (d) compared with a $\mu$SR experiment, $\lambda_{ab}(0) = 4,260\,\AA$.~\cite{par10} Fitting to a BCS formula we found the superconducting gap to be $3.0 \pm 0.3$ meV consistent with Knight shift data.}
\label{fig5}
\end{figure}

We next turn  our discussion to the diamagnetism and vortex supercurrents in the superconducting state. The linewidths of the central transitions of $^{75}$As and $^{23}$Na are shown in Fig.4. Below $T_c$ the linewidths from both samples broaden due to vortices and approach a constant at low temperature approximately at the minimum in $^{75}T_2$, which we identify with the formation of a robust vortex solid.  In our earlier work on Co doped Ba(Fe$_{0.926}$Co$_{0.074}$)$_2$As$_2$~\cite{oh11} we  have associated a minimum $T_2$ with the irreversibility temperature where vortices become pinned.  Although $^{23}$Na NMR is not as effective as  $^{75}$As NMR for probing the electronic excitations, it nonetheless provides a higher resolution tool for analysis of the spectrum in the superconducting state which is affected by vortices and from which we can determine the penetration depth, $\lambda_{ab}$.  The comparisons between the superconducting state spectra and the normal state spectra  in Fig.5,  indicate an increase in the high frequency portion of the spectrum as expected in a vortex solid.  We have solved the Ginzburg-Landau (GL) equations in an ideal vortex lattice using Brandt's algorithm,~\cite{bra97} where we set $H_{c2}$ = 36 T.~\cite{zho13} The resulting best fit distribution of local fields for $^{23}$Na at $T = 4.2$ K is presented in Fig.5(c). The fit was performed by convoluting the GL solution from the algorithm with the normal state ($T = 20$ K) spectrum, minimizing its difference from the experimental spectra using the penetration depth as an adjustable parameter. We have analyzed $\lambda_{ab}(T)$ using a BCS  formula in the low temperature limit, $\lambda(T) = \lambda(0)[1+\sqrt{\frac{\pi\Delta}{2k_BT}}\exp(-\frac{\Delta}{k_BT})]$.~\cite{pro06} For  $^{75}$As we find $\lambda_{ab}$(0) to be $4,780\,\AA$ and from $^{23}$Na, $\lambda_{ab}(0) = 5,327\,\AA$, which can be compared with $\mu$SR,~\cite{par10} $\lambda_{ab}(0) = 4,260\,\AA$. The penetration depth from the $^{23}$Na is more reliable since the temperature dependent spin shift is negligible.    From the fit of the temperature dependent penetration depth to the BCS formula and Eq.3 and 4 we found the superconducting gap size $3.0 \pm 0.4$ meV, close to the value from the Knight shift, supporting the view that  NaCo25 superconductivity is close to weak-coupling.

\section{CONCLUSION} 
In summary, we have found evidence for weak-coupling superconductivity from the temperature dependence of the Knight shift and the penetration depth in NaCo25 consistent with BCS theory. The superconducting gap size taken from the Knight shift is $3.6 \pm 0.3$ meV and $3.0 \pm 0.4$ meV from the penetration depth.  From the Knight shift we have $2\Delta/k_BT_c = 4.0 \pm 0.3$  close to the weak-coupling BCS limit of 3.53.  From Ginzburg-Landau theory we find a zero temperature penetration depth of $\lambda_{ab}(0) = 5,327 \pm$ 78$\,\AA$.

\section{Acknowledgments}
Research was supported by the U.S. Department of Energy, Office of Basic Energy Sciences, Division of Materials Sciences and Engineering under Awards DE-FG02-05ER46248 (Northwestern University). The single crystal growth at the University of Tennessee was supported by U.S. DOE, BES under grant No, DE-FG02-05ER46202(P.D.)\\

\begin{thebibliography}{23}%
\makeatletter
\providecommand \@ifxundefined [1]{%
 \@ifx{#1\undefined}
}%
\providecommand \@ifnum [1]{%
 \ifnum #1\expandafter \@firstoftwo
 \else \expandafter \@secondoftwo
 \fi
}%
\providecommand \@ifx [1]{%
 \ifx #1\expandafter \@firstoftwo
 \else \expandafter \@secondoftwo
 \fi
}%
\providecommand \natexlab [1]{#1}%
\providecommand \enquote  [1]{``#1''}%
\providecommand \bibnamefont  [1]{#1}%
\providecommand \bibfnamefont [1]{#1}%
\providecommand \citenamefont [1]{#1}%
\providecommand \href@noop [0]{\@secondoftwo}%
\providecommand \href [0]{\begingroup \@sanitize@url \@href}%
\providecommand \@href[1]{\@@startlink{#1}\@@href}%
\providecommand \@@href[1]{\endgroup#1\@@endlink}%
\providecommand \@sanitize@url [0]{\catcode `\\12\catcode `\$12\catcode
  `\&12\catcode `\#12\catcode `\^12\catcode `\_12\catcode `\%12\relax}%
\providecommand \@@startlink[1]{}%
\providecommand \@@endlink[0]{}%
\providecommand \url  [0]{\begingroup\@sanitize@url \@url }%
\providecommand \@url [1]{\endgroup\@href {#1}{\urlprefix }}%
\providecommand \urlprefix  [0]{URL }%
\providecommand \Eprint [0]{\href }%
\providecommand \doibase [0]{http://dx.doi.org/}%
\providecommand \selectlanguage [0]{\@gobble}%
\providecommand \bibinfo  [0]{\@secondoftwo}%
\providecommand \bibfield  [0]{\@secondoftwo}%
\providecommand \translation [1]{[#1]}%
\providecommand \BibitemOpen [0]{}%
\providecommand \bibitemStop [0]{}%
\providecommand \bibitemNoStop [0]{.\EOS\space}%
\providecommand \EOS [0]{\spacefactor3000\relax}%
\providecommand \BibitemShut  [1]{\csname bibitem#1\endcsname}%
\let\auto@bib@innerbib\@empty
\bibitem [{\citenamefont {Norman}(2013)}]{nor13}%
  \BibitemOpen
  \bibfield  {author} {\bibinfo {author} {\bibfnamefont {M.~R.}\ \bibnamefont
  {Norman}},\ }\href@noop {} {\  (\bibinfo {year} {2013})},\ \Eprint
  {http://arxiv.org/abs/arXiv.org:1302.3176v1} {arXiv.org:1302.3176v1}
  \BibitemShut {NoStop}%
\bibitem [{\citenamefont {Bardeen}\ \emph {et~al.}(1957)\citenamefont
  {Bardeen}, \citenamefont {Cooper},\ and\ \citenamefont {Schrieffer}}]{bar57}%
  \BibitemOpen
  \bibfield  {author} {\bibinfo {author} {\bibfnamefont {J.}~\bibnamefont
  {Bardeen}}, \bibinfo {author} {\bibfnamefont {L.}~\bibnamefont {Cooper}}, \
  and\ \bibinfo {author} {\bibfnamefont {J.~R.}\ \bibnamefont {Schrieffer}},\
  }\href@noop {} {\bibfield  {journal} {\bibinfo  {journal} {Phys. Rev.}\
  }\textbf {\bibinfo {volume} {108}},\ \bibinfo {pages} {1175} (\bibinfo {year}
  {1957})}\BibitemShut {NoStop}%
\bibitem [{\citenamefont {Lee}(1996)}]{lee96}%
  \BibitemOpen
  \bibfield  {author} {\bibinfo {author} {\bibfnamefont {D.~M.}\ \bibnamefont
  {Lee}},\ }\href@noop {} {\bibfield  {journal} {\bibinfo  {journal} {Rev. Mod.
  Phys.}\ }\textbf {\bibinfo {volume} {69}},\ \bibinfo {pages} {645} (\bibinfo
  {year} {1996})}\BibitemShut {NoStop}%
\bibitem [{\citenamefont {Liu}\ \emph {et~al.}(2011)\citenamefont {Liu},
  \citenamefont {Richard}, \citenamefont {Nakayama}, \citenamefont {Chen},
  \citenamefont {Dong}, \citenamefont {He}, \citenamefont {Wang}, \citenamefont
  {Xia}, \citenamefont {Umezawa}, \citenamefont {Kawahara}, \citenamefont
  {Souma}, \citenamefont {Sato}, \citenamefont {Takahashi}, \citenamefont
  {Qian}, \citenamefont {Huang}, \citenamefont {Xu}, \citenamefont {Shi},
  \citenamefont {Ding},\ and\ \citenamefont {Wang}}]{liu11}%
  \BibitemOpen
  \bibfield  {author} {\bibinfo {author} {\bibfnamefont {Z.~H.}\ \bibnamefont
  {Liu}}, \bibinfo {author} {\bibfnamefont {P.}~\bibnamefont {Richard}},
  \bibinfo {author} {\bibfnamefont {K.}~\bibnamefont {Nakayama}}, \bibinfo
  {author} {\bibfnamefont {G.~F.}\ \bibnamefont {Chen}}, \bibinfo {author}
  {\bibfnamefont {S.}~\bibnamefont {Dong}}, \bibinfo {author} {\bibfnamefont
  {J.~B.}\ \bibnamefont {He}}, \bibinfo {author} {\bibfnamefont {D.~M.}\
  \bibnamefont {Wang}}, \bibinfo {author} {\bibfnamefont {T.~L.}\ \bibnamefont
  {Xia}}, \bibinfo {author} {\bibfnamefont {K.}~\bibnamefont {Umezawa}},
  \bibinfo {author} {\bibfnamefont {T.}~\bibnamefont {Kawahara}}, \bibinfo
  {author} {\bibfnamefont {S.}~\bibnamefont {Souma}}, \bibinfo {author}
  {\bibfnamefont {T.}~\bibnamefont {Sato}}, \bibinfo {author} {\bibfnamefont
  {T.}~\bibnamefont {Takahashi}}, \bibinfo {author} {\bibfnamefont
  {T.}~\bibnamefont {Qian}}, \bibinfo {author} {\bibfnamefont {Y.}~\bibnamefont
  {Huang}}, \bibinfo {author} {\bibfnamefont {N.}~\bibnamefont {Xu}}, \bibinfo
  {author} {\bibfnamefont {Y.}~\bibnamefont {Shi}}, \bibinfo {author}
  {\bibfnamefont {H.}~\bibnamefont {Ding}}, \ and\ \bibinfo {author}
  {\bibfnamefont {S.~C.}\ \bibnamefont {Wang}},\ }\href@noop {} {\bibfield
  {journal} {\bibinfo  {journal} {Phys. Rev. B.}\ }\textbf {\bibinfo {volume}
  {84}},\ \bibinfo {pages} {064519} (\bibinfo {year} {2011})}\BibitemShut
  {NoStop}%
\bibitem [{\citenamefont {Thirupathaiah}\ \emph {et~al.}(2012)\citenamefont
  {Thirupathaiah}, \citenamefont {Evtushinsky}, \citenamefont {Maletz},
  \citenamefont {Zabolotnyy}, \citenamefont {Kordyuk}, \citenamefont {Kim},
  \citenamefont {Wurmehl}, \citenamefont {Roslova}, \citenamefont {Morozov},
  \citenamefont {B\"uchner},\ and\ \citenamefont {Borisenko}}]{thi12}%
  \BibitemOpen
  \bibfield  {author} {\bibinfo {author} {\bibfnamefont {S.}~\bibnamefont
  {Thirupathaiah}}, \bibinfo {author} {\bibfnamefont {D.~V.}\ \bibnamefont
  {Evtushinsky}}, \bibinfo {author} {\bibfnamefont {J.}~\bibnamefont {Maletz}},
  \bibinfo {author} {\bibfnamefont {V.~B.}\ \bibnamefont {Zabolotnyy}},
  \bibinfo {author} {\bibfnamefont {A.~A.}\ \bibnamefont {Kordyuk}}, \bibinfo
  {author} {\bibfnamefont {T.~K.}\ \bibnamefont {Kim}}, \bibinfo {author}
  {\bibfnamefont {S.}~\bibnamefont {Wurmehl}}, \bibinfo {author} {\bibfnamefont
  {M.}~\bibnamefont {Roslova}}, \bibinfo {author} {\bibfnamefont
  {I.}~\bibnamefont {Morozov}}, \bibinfo {author} {\bibfnamefont
  {B.}~\bibnamefont {B\"uchner}}, \ and\ \bibinfo {author} {\bibfnamefont
  {S.~V.}\ \bibnamefont {Borisenko}},\ }\href@noop {} {\bibfield  {journal}
  {\bibinfo  {journal} {Phys. Rev. B}\ }\textbf {\bibinfo {volume} {86}},\
  \bibinfo {pages} {214508} (\bibinfo {year} {2012})}\BibitemShut {NoStop}%
\bibitem [{\citenamefont {Cai}\ \emph {et~al.}(2013)\citenamefont {Cai},
  \citenamefont {Zhou}, \citenamefont {Ruan}, \citenamefont {Wang},
  \citenamefont {Chen}, \citenamefont {Lee},\ and\ \citenamefont
  {Wang}}]{cai13}%
  \BibitemOpen
  \bibfield  {author} {\bibinfo {author} {\bibfnamefont {P.}~\bibnamefont
  {Cai}}, \bibinfo {author} {\bibfnamefont {X.}~\bibnamefont {Zhou}}, \bibinfo
  {author} {\bibfnamefont {W.}~\bibnamefont {Ruan}}, \bibinfo {author}
  {\bibfnamefont {A.}~\bibnamefont {Wang}}, \bibinfo {author} {\bibfnamefont
  {X.}~\bibnamefont {Chen}}, \bibinfo {author} {\bibfnamefont {D.~H.}\
  \bibnamefont {Lee}}, \ and\ \bibinfo {author} {\bibfnamefont
  {Y.}~\bibnamefont {Wang}},\ }\href@noop {} {\bibfield  {journal} {\bibinfo
  {journal} {Nauture Comm.}\ }\textbf {\bibinfo {volume} {4}},\ \bibinfo
  {pages} {1596} (\bibinfo {year} {2013})}\BibitemShut {NoStop}%
\bibitem [{\citenamefont {Yang}\ \emph {et~al.}(2012)\citenamefont {Yang},
  \citenamefont {Wang}, \citenamefont {Fang}, \citenamefont {Li}, \citenamefont
  {Kariyado}, \citenamefont {Chen}, \citenamefont {Ogata}, \citenamefont {Das},
  \citenamefont {Balatsky},\ and\ \citenamefont {Wen}}]{yan12}%
  \BibitemOpen
  \bibfield  {author} {\bibinfo {author} {\bibfnamefont {H.}~\bibnamefont
  {Yang}}, \bibinfo {author} {\bibfnamefont {Z.}~\bibnamefont {Wang}}, \bibinfo
  {author} {\bibfnamefont {D.}~\bibnamefont {Fang}}, \bibinfo {author}
  {\bibfnamefont {S.}~\bibnamefont {Li}}, \bibinfo {author} {\bibfnamefont
  {T.}~\bibnamefont {Kariyado}}, \bibinfo {author} {\bibfnamefont
  {G.}~\bibnamefont {Chen}}, \bibinfo {author} {\bibfnamefont {M.}~\bibnamefont
  {Ogata}}, \bibinfo {author} {\bibfnamefont {T.}~\bibnamefont {Das}}, \bibinfo
  {author} {\bibfnamefont {A.~V.}\ \bibnamefont {Balatsky}}, \ and\ \bibinfo
  {author} {\bibfnamefont {H.~H.}\ \bibnamefont {Wen}},\ }\href@noop {}
  {\bibfield  {journal} {\bibinfo  {journal} {Phys. Rev. B.}\ }\textbf
  {\bibinfo {volume} {86}},\ \bibinfo {pages} {214512} (\bibinfo {year}
  {2012})}\BibitemShut {NoStop}%
\bibitem [{\citenamefont {Wang}\ \emph {et~al.}(2013)\citenamefont {Wang},
  \citenamefont {Yang}, \citenamefont {Fang}, \citenamefont {Shen},
  \citenamefont {Wang}, \citenamefont {Shan}, \citenamefont {Zhang},
  \citenamefont {Dai},\ and\ \citenamefont {Wen}}]{wan13}%
  \BibitemOpen
  \bibfield  {author} {\bibinfo {author} {\bibfnamefont {Z.}~\bibnamefont
  {Wang}}, \bibinfo {author} {\bibfnamefont {H.}~\bibnamefont {Yang}}, \bibinfo
  {author} {\bibfnamefont {D.}~\bibnamefont {Fang}}, \bibinfo {author}
  {\bibfnamefont {B.}~\bibnamefont {Shen}}, \bibinfo {author} {\bibfnamefont
  {Q.}~\bibnamefont {Wang}}, \bibinfo {author} {\bibfnamefont {L.}~\bibnamefont
  {Shan}}, \bibinfo {author} {\bibfnamefont {C.}~\bibnamefont {Zhang}},
  \bibinfo {author} {\bibfnamefont {P.}~\bibnamefont {Dai}}, \ and\ \bibinfo
  {author} {\bibfnamefont {H.~H.}\ \bibnamefont {Wen}},\ }\href@noop {}
  {\bibfield  {journal} {\bibinfo  {journal} {Nature Physics}\ }\textbf
  {\bibinfo {volume} {9}},\ \bibinfo {pages} {42} (\bibinfo {year}
  {2013})}\BibitemShut {NoStop}%
\bibitem [{\citenamefont {Mitrovi\'c}\ \emph {et~al.}(2001)\citenamefont
  {Mitrovi\'c}, \citenamefont {Sigmund},\ and\ \citenamefont
  {Halperin}}]{mit01a}%
  \BibitemOpen
  \bibfield  {author} {\bibinfo {author} {\bibfnamefont {V.~F.}\ \bibnamefont
  {Mitrovi\'c}}, \bibinfo {author} {\bibfnamefont {E.~E.}\ \bibnamefont
  {Sigmund}}, \ and\ \bibinfo {author} {\bibfnamefont {W.~P.}\ \bibnamefont
  {Halperin}},\ }\href@noop {} {\bibfield  {journal} {\bibinfo  {journal}
  {Phys. Rev. B}\ }\textbf {\bibinfo {volume} {64}},\ \bibinfo {pages} {024520}
  (\bibinfo {year} {2001})}\BibitemShut {NoStop}%
\bibitem [{\citenamefont {Oh}\ \emph {et~al.}(2011)\citenamefont {Oh},
  \citenamefont {Mounce}, \citenamefont {Mukhopadhyay}, \citenamefont
  {Halperin}, \citenamefont {Vorontsov}, \citenamefont {Bud'ko}, \citenamefont
  {Canfield}, \citenamefont {Furukawa}, \citenamefont {Reyes},\ and\
  \citenamefont {Kuhns}}]{oh11}%
  \BibitemOpen
  \bibfield  {author} {\bibinfo {author} {\bibfnamefont {S.}~\bibnamefont
  {Oh}}, \bibinfo {author} {\bibfnamefont {A.~M.}\ \bibnamefont {Mounce}},
  \bibinfo {author} {\bibfnamefont {S.}~\bibnamefont {Mukhopadhyay}}, \bibinfo
  {author} {\bibfnamefont {W.~P.}\ \bibnamefont {Halperin}}, \bibinfo {author}
  {\bibfnamefont {A.~B.}\ \bibnamefont {Vorontsov}}, \bibinfo {author}
  {\bibfnamefont {S.~L.}\ \bibnamefont {Bud'ko}}, \bibinfo {author}
  {\bibfnamefont {P.~C.}\ \bibnamefont {Canfield}}, \bibinfo {author}
  {\bibfnamefont {Y.}~\bibnamefont {Furukawa}}, \bibinfo {author}
  {\bibfnamefont {A.~P.}\ \bibnamefont {Reyes}}, \ and\ \bibinfo {author}
  {\bibfnamefont {P.~L.}\ \bibnamefont {Kuhns}},\ }\href@noop {} {\bibfield
  {journal} {\bibinfo  {journal} {Phys. Rev. B}\ }\textbf {\bibinfo {volume}
  {83}},\ \bibinfo {pages} {214501} (\bibinfo {year} {2011})}\BibitemShut
  {NoStop}%
\bibitem [{\citenamefont {Oh}\ \emph {et~al.}(2012)\citenamefont {Oh},
  \citenamefont {Mounce}, \citenamefont {Halperin}, \citenamefont {Zhang},
  \citenamefont {Dai}, \citenamefont {Reyes},\ and\ \citenamefont
  {Kuhns}}]{oh12}%
  \BibitemOpen
  \bibfield  {author} {\bibinfo {author} {\bibfnamefont {S.}~\bibnamefont
  {Oh}}, \bibinfo {author} {\bibfnamefont {A.~M.}\ \bibnamefont {Mounce}},
  \bibinfo {author} {\bibfnamefont {W.~P.}\ \bibnamefont {Halperin}}, \bibinfo
  {author} {\bibfnamefont {C.~L.}\ \bibnamefont {Zhang}}, \bibinfo {author}
  {\bibfnamefont {P.}~\bibnamefont {Dai}}, \bibinfo {author} {\bibfnamefont
  {A.~P.}\ \bibnamefont {Reyes}}, \ and\ \bibinfo {author} {\bibfnamefont
  {P.~L.}\ \bibnamefont {Kuhns}},\ }\href@noop {} {\bibfield  {journal}
  {\bibinfo  {journal} {Phys. Rev. B}\ }\textbf {\bibinfo {volume} {85}},\
  \bibinfo {pages} {174508} (\bibinfo {year} {2012})}\BibitemShut {NoStop}%
\bibitem [{\citenamefont {Ning}\ \emph {et~al.}(2010)\citenamefont {Ning},
  \citenamefont {Ahilan}, \citenamefont {Imai}, \citenamefont {Sefat},
  \citenamefont {McGuire}, \citenamefont {Sales}, \citenamefont {Mandrus},
  \citenamefont {Cheng}, \citenamefont {Shen},\ and\ \citenamefont
  {Wen}}]{nin10}%
  \BibitemOpen
  \bibfield  {author} {\bibinfo {author} {\bibfnamefont {F.~L.}\ \bibnamefont
  {Ning}}, \bibinfo {author} {\bibfnamefont {K.}~\bibnamefont {Ahilan}},
  \bibinfo {author} {\bibfnamefont {T.}~\bibnamefont {Imai}}, \bibinfo {author}
  {\bibfnamefont {A.~S.}\ \bibnamefont {Sefat}}, \bibinfo {author}
  {\bibfnamefont {M.~A.}\ \bibnamefont {McGuire}}, \bibinfo {author}
  {\bibfnamefont {B.~C.}\ \bibnamefont {Sales}}, \bibinfo {author}
  {\bibfnamefont {D.}~\bibnamefont {Mandrus}}, \bibinfo {author} {\bibfnamefont
  {P.}~\bibnamefont {Cheng}}, \bibinfo {author} {\bibfnamefont
  {B.}~\bibnamefont {Shen}}, \ and\ \bibinfo {author} {\bibfnamefont {H.~H.}\
  \bibnamefont {Wen}},\ }\href@noop {} {\bibfield  {journal} {\bibinfo
  {journal} {Phys. Rev. Lett.}\ }\textbf {\bibinfo {volume} {104}},\ \bibinfo
  {pages} {037001} (\bibinfo {year} {2010})}\BibitemShut {NoStop}%
\bibitem [{\citenamefont {Yosida}(1958)}]{yos58}%
  \BibitemOpen
  \bibfield  {author} {\bibinfo {author} {\bibfnamefont {K.}~\bibnamefont
  {Yosida}},\ }\href@noop {} {\bibfield  {journal} {\bibinfo  {journal} {Phys.
  Rev.}\ }\textbf {\bibinfo {volume} {110}},\ \bibinfo {pages} {769} (\bibinfo
  {year} {1958})}\BibitemShut {NoStop}%
\bibitem [{\citenamefont {Padamsee}\ \emph {et~al.}(1973)\citenamefont
  {Padamsee}, \citenamefont {Neighbor},\ and\ \citenamefont
  {Shiffman}}]{pad73}%
  \BibitemOpen
  \bibfield  {author} {\bibinfo {author} {\bibfnamefont {H.}~\bibnamefont
  {Padamsee}}, \bibinfo {author} {\bibfnamefont {J.~E.}\ \bibnamefont
  {Neighbor}}, \ and\ \bibinfo {author} {\bibfnamefont {C.~A.}\ \bibnamefont
  {Shiffman}},\ }\href@noop {} {\bibfield  {journal} {\bibinfo  {journal} {J.
  Low Temp. Phys.}\ }\textbf {\bibinfo {volume} {12}},\ \bibinfo {pages} {387}
  (\bibinfo {year} {1973})}\BibitemShut {NoStop}%
\bibitem [{\citenamefont {Halperin}\ and\ \citenamefont
  {Varoquaux}(1990)}]{hal90}%
  \BibitemOpen
  \bibfield  {author} {\bibinfo {author} {\bibfnamefont {W.~P.}\ \bibnamefont
  {Halperin}}\ and\ \bibinfo {author} {\bibfnamefont {E.}~\bibnamefont
  {Varoquaux}},\ }in\ \href@noop {} {\emph {\bibinfo {booktitle} {Helium
  Three}}},\ \bibinfo {series} {{Modern Problems in Condensed Matter
  Sciences}}, Vol.~\bibinfo {volume} {26},\ \bibinfo {editor} {edited by\
  \bibinfo {editor} {\bibfnamefont {W.~P.}\ \bibnamefont {Halperin}}\ and\
  \bibinfo {editor} {\bibfnamefont {L.~P.}\ \bibnamefont {Pitaevskii}}}\
  (\bibinfo  {publisher} {Elsevier},\ \bibinfo {year} {1990})\ Chap.~\bibinfo
  {chapter} {7}, p.\ \bibinfo {pages} {353}\BibitemShut {NoStop}%
\bibitem [{\citenamefont {Sauls}()}]{sau12}%
  \BibitemOpen
  \bibfield  {author} {\bibinfo {author} {\bibfnamefont {J.~A.}\ \bibnamefont
  {Sauls}},\ }\href@noop {} {}\bibinfo {note} {Private
  communication}\BibitemShut {NoStop}%
\bibitem [{\citenamefont {Kitagawa}\ \emph {et~al.}(2011)\citenamefont
  {Kitagawa}, \citenamefont {Mezaki}, \citenamefont {Matsubayashi},
  \citenamefont {Uwatoko},\ and\ \citenamefont {Takigawa}}]{kit11}%
  \BibitemOpen
  \bibfield  {author} {\bibinfo {author} {\bibfnamefont {K.}~\bibnamefont
  {Kitagawa}}, \bibinfo {author} {\bibfnamefont {Y.}~\bibnamefont {Mezaki}},
  \bibinfo {author} {\bibfnamefont {K.}~\bibnamefont {Matsubayashi}}, \bibinfo
  {author} {\bibfnamefont {Y.}~\bibnamefont {Uwatoko}}, \ and\ \bibinfo
  {author} {\bibfnamefont {M.}~\bibnamefont {Takigawa}},\ }\href@noop {}
  {\bibfield  {journal} {\bibinfo  {journal} {J. Phys. Soc. Jpn.}\ }\textbf
  {\bibinfo {volume} {80}},\ \bibinfo {pages} {033705} (\bibinfo {year}
  {2011})}\BibitemShut {NoStop}%
\bibitem [{\citenamefont {Li}\ \emph {et~al.}(2011)\citenamefont {Li},
  \citenamefont {Sun}, \citenamefont {Lin}, \citenamefont {Su}, \citenamefont
  {Hu},\ and\ \citenamefont {Zheng}}]{li11}%
  \BibitemOpen
  \bibfield  {author} {\bibinfo {author} {\bibfnamefont {Z.}~\bibnamefont
  {Li}}, \bibinfo {author} {\bibfnamefont {D.~L.}\ \bibnamefont {Sun}},
  \bibinfo {author} {\bibfnamefont {C.~T.}\ \bibnamefont {Lin}}, \bibinfo
  {author} {\bibfnamefont {Y.~H.}\ \bibnamefont {Su}}, \bibinfo {author}
  {\bibfnamefont {J.~P.}\ \bibnamefont {Hu}}, \ and\ \bibinfo {author}
  {\bibfnamefont {G.~Q.}\ \bibnamefont {Zheng}},\ }\href@noop {} {\bibfield
  {journal} {\bibinfo  {journal} {Phys. Rev. B}\ }\textbf {\bibinfo {volume}
  {83}},\ \bibinfo {pages} {140506} (\bibinfo {year} {2011})}\BibitemShut
  {NoStop}%
\bibitem [{\citenamefont {Moriya}(1963)}]{mor63}%
  \BibitemOpen
  \bibfield  {author} {\bibinfo {author} {\bibfnamefont {T.}~\bibnamefont
  {Moriya}},\ }\href@noop {} {\bibfield  {journal} {\bibinfo  {journal} {J.
  Phys. Soc. Jpn.}\ }\textbf {\bibinfo {volume} {77}},\ \bibinfo {pages}
  {114709} (\bibinfo {year} {1963})}\BibitemShut {NoStop}%
\bibitem [{\citenamefont {Parker}\ \emph {et~al.}(2010)\citenamefont {Parker},
  \citenamefont {Smith}, \citenamefont {Lancaster}, \citenamefont {Steele},
  \citenamefont {Franke}, \citenamefont {Baker}, \citenamefont {Pratt},
  \citenamefont {Pitcher}, \citenamefont {Blundell},\ and\ \citenamefont
  {Clarke}}]{par10}%
  \BibitemOpen
  \bibfield  {author} {\bibinfo {author} {\bibfnamefont {D.~R.}\ \bibnamefont
  {Parker}}, \bibinfo {author} {\bibfnamefont {M.~J.~P.}\ \bibnamefont
  {Smith}}, \bibinfo {author} {\bibfnamefont {T.}~\bibnamefont {Lancaster}},
  \bibinfo {author} {\bibfnamefont {A.~J.}\ \bibnamefont {Steele}}, \bibinfo
  {author} {\bibfnamefont {I.}~\bibnamefont {Franke}}, \bibinfo {author}
  {\bibfnamefont {P.~J.}\ \bibnamefont {Baker}}, \bibinfo {author}
  {\bibfnamefont {F.~L.}\ \bibnamefont {Pratt}}, \bibinfo {author}
  {\bibfnamefont {M.~J.}\ \bibnamefont {Pitcher}}, \bibinfo {author}
  {\bibfnamefont {S.~J.}\ \bibnamefont {Blundell}}, \ and\ \bibinfo {author}
  {\bibfnamefont {S.~J.}\ \bibnamefont {Clarke}},\ }\href@noop {} {\bibfield
  {journal} {\bibinfo  {journal} {Phys. Rev. Lett.}\ }\textbf {\bibinfo
  {volume} {104}},\ \bibinfo {pages} {057007} (\bibinfo {year}
  {2010})}\BibitemShut {NoStop}%
\bibitem [{\citenamefont {Brandt}(1997)}]{bra97}%
  \BibitemOpen
  \bibfield  {author} {\bibinfo {author} {\bibfnamefont {E.~H.}\ \bibnamefont
  {Brandt}},\ }\href@noop {} {\bibfield  {journal} {\bibinfo  {journal} {Phys.
  Rev. Lett.}\ }\textbf {\bibinfo {volume} {78}},\ \bibinfo {pages} {2208}
  (\bibinfo {year} {1997})}\BibitemShut {NoStop}%
\bibitem [{\citenamefont {Zhou}\ \emph {et~al.}(2013)\citenamefont {Zhou},
  \citenamefont {Hong}, \citenamefont {Qiu}, \citenamefont {Pan}, \citenamefont
  {Zhang}, \citenamefont {Li}, \citenamefont {Dong}, \citenamefont {Wang},
  \citenamefont {Luo}, \citenamefont {Chen},\ and\ \citenamefont {Li}}]{zho13}%
  \BibitemOpen
  \bibfield  {author} {\bibinfo {author} {\bibfnamefont {S.~Y.}\ \bibnamefont
  {Zhou}}, \bibinfo {author} {\bibfnamefont {X.~C.}\ \bibnamefont {Hong}},
  \bibinfo {author} {\bibfnamefont {x.}~\bibnamefont {Qiu}}, \bibinfo {author}
  {\bibfnamefont {B.~Y.}\ \bibnamefont {Pan}}, \bibinfo {author} {\bibfnamefont
  {Z.}~\bibnamefont {Zhang}}, \bibinfo {author} {\bibfnamefont {X.~L.}\
  \bibnamefont {Li}}, \bibinfo {author} {\bibfnamefont {W.~N.}\ \bibnamefont
  {Dong}}, \bibinfo {author} {\bibfnamefont {A.~F.}\ \bibnamefont {Wang}},
  \bibinfo {author} {\bibfnamefont {X.~G.}\ \bibnamefont {Luo}}, \bibinfo
  {author} {\bibfnamefont {X.~H.}\ \bibnamefont {Chen}}, \ and\ \bibinfo
  {author} {\bibfnamefont {S.~Y.}\ \bibnamefont {Li}},\ }\href@noop {}
  {\bibfield  {journal} {\bibinfo  {journal} {EPL.}\ }\textbf {\bibinfo
  {volume} {101}},\ \bibinfo {pages} {17007} (\bibinfo {year}
  {2013})}\BibitemShut {NoStop}%
\bibitem [{\citenamefont {Prozorov}\ and\ \citenamefont
  {Giannetta}(2006)}]{pro06}%
  \BibitemOpen
  \bibfield  {author} {\bibinfo {author} {\bibfnamefont {R.}~\bibnamefont
  {Prozorov}}\ and\ \bibinfo {author} {\bibfnamefont {W.}~\bibnamefont
  {Giannetta}},\ }\href@noop {} {\bibfield  {journal} {\bibinfo  {journal}
  {Supercond. Sci. Technol.}\ }\textbf {\bibinfo {volume} {19}},\ \bibinfo
  {pages} {41} (\bibinfo {year} {2006})}\BibitemShut {NoStop}%
\end{thebibliography}

%

\end{document}